# Understanding State-level Variations in U.S. Infant Mortality: 2000–2015

Authors: Alireza Ebrahimvandi, Niyousha Hosseinichimeh, Jay Iams

**Objective**: To exploit state variations in infant mortality over time, identify diagnoses that contributed to reduction of the infant mortality rate (IMR) between 2000 and 2015, and examine factors associated with preterm related deaths at the state level.

**Methods**: Using linked birth-infant deaths period data files from 2000 to 2015, we examined patterns in the leading causes of infant deaths including (1) the preterm-related mortality rate (PMR), (2) the congenital malformation-related mortality rate (CMR), (3) the sudden infant death syndrome mortality rate (SMR), and (4) all other 130 major mortality causes (OMR). We compared these rates at both national and state levels in 2000 and 2015 to find reduction trends. Creating a cross-sectional time series of states' PMR data and some explanatory variables, we implemented a fixed-effect regression model controlling for infant, maternal, and institutional characteristics.

**Results**: We found substantial state-level variations in changes of the IMR (range= -2.87 to 2.08), PMR (range= -1.77 to 0.67) and the CMR (range= -1.23 to 0.59) between 2000 and 2015. Twenty-one states in which the IMR declined more than the national average of 0.99 (from 6.89 to 5.90) were labeled as *successful*. We also labeled 20 states that saw a decline in their IMR less than the national average as *unsuccessful*. In the successful states, we found a reduction in the PMR accounted for the largest decline in the IMR—0.90 fewer deaths per 1,000 live births—or six times more than the PMR decline (0.14) in unsuccessful states. Changes in the other subgroups did not differ significantly in successful and unsuccessful states. Regression results showed that the PMR is positively associated with inadequate care (P-Value <0.05). A one-percentage-point decline in the share of pregnant women with inadequate care is significantly associated with 0.011 fewer preterm-related deaths per 1,000 live births. The magnitude of this variable is small relative to the PMR mean. The percentage of teen pregnancies, multiple births, and pregnant women that smoke was not significantly associated with the state-level PMR.

*Policy Implications*: Trends in the leading causes of mortality reduction are heterogeneous across states. States with high infant mortality need to focus on preterm-related deaths, as they are the largest contributor to the success of states with a high infant-death reduction. Although its impact is not large, reducing the percentage of pregnant women with inadequate care is one of the mechanisms through which preterm-related deaths might decrease.

## Background

The U.S. infant mortality rate (IMR) increased in 2015 after a decade of decline—from 2005 to 2014 (1). The U.S. IMR is 71% higher than the average rate for comparable countries in the Organization for Economic Co-operation and Development (OECD) (2). Although the U.S. rate declined by 14%, from 6.89 in 2000 to 5.90 in 2015, infant mortality has declined more slowly than in comparable countries. The IMR is defined as the number of infant deaths per 1,000 live births before their first birthday, and it is a representation of population health, quality of health care, and societal well-being (3).

Preterm birth, one of the most complicated factors associated with infant mortality (4, 5), has become a focus of research. Despite the reduction in recent decades, preterm birth in the U.S. (9.8%) remains higher than in European countries such as France (7.5%), Germany (8.3%), and Sweden (5.8%) (6). States vary substantially in terms of IMR size, trends, and preterm-related mortality rate (PMR). Figure 1.a shows the rate of infant mortality for 50 states from 2000 to 2015. The size of each dot is proportionate to the number of births in corresponding states and years. The continuous line shows the national average at each year. Four observations can be inferred from Figure 1.a: (1) states differ considerably in their IMR each year, (2) the IMR in larger states tend to be closer to the average, (3) states experienced a diverse change in their IMR from 2000 to 2015, and (4) on average, states experienced a decline in their IMR. However, the rates in Maine, South Dakota, and Texas are all higher in 2015 than in 2000. States also vary in mortality rates for babies born preterm. Figure 1.b shows the changes in PMR between 2000 and 2015 for each state. States such as Tennessee and Michigan reduced the PMR by more than 1.2 deaths per 1,000 live births, while the PMR increased by more than 0.5 deaths per 1,000 live births in Alaska, Maine, and South Dakota.

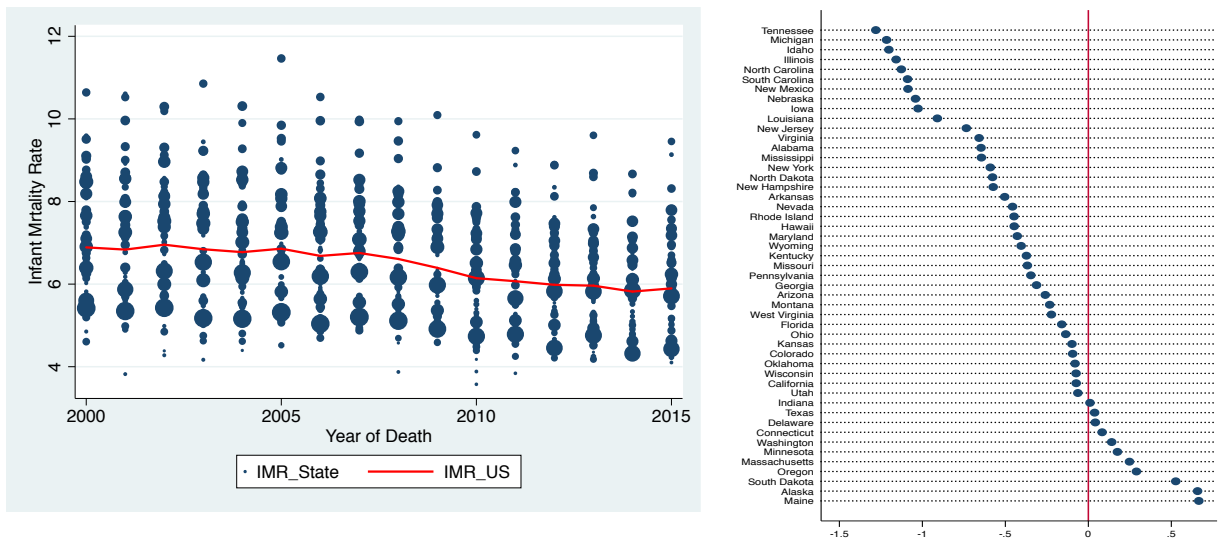

a. Infant mortality for each state from 2000 to 2015.

b. PMR reductions: per 1000 live births, between 2000 and 2015.

Figure 1. Trends of states' infant mortality and preterm-related mortality rates from 2000-2015.[1]

Note: The District of Columbia was dropped since the large change is attributed to demographic changes over the past decade.

The Centers for Disease Control and Prevention (CDC) reports national trends of infant mortality based on five leading causes that include: congenital malformations and chromosomal abnormalities, disorders related to short gestation and low birth weight, newborns affected by maternal complications of pregnancy, sudden infant death syndrome, and accidents. Change in state IMRs and variation across states have been reported from 2005 to 2014 (7), and infant mortality trends were also investigated in limited geographical locations. In five Southeastern states with the highest IMRs in the U.S., descriptive IMR statistics based on location, causes of death, infant and maternal characteristics were reported from 2005 to 2009 (8). Finally, a study of perinatal mortality in 50 states between 1989 and 2000 found that perinatal mortality among whites increased because of an increase in medically indicated preterm births (9). However, there are no studies comparing PMR trends by states over time. Many studies have focused on one state or a limited number of modifiable risk factors across states to investigate determinants of preterm births (9-13).

Understanding drivers of IMR reduction in states that have achieved substantial rate reductions may help to improve high IMRs. In this study, we explore state variations of the IMR and categorize them as either successful—with an IMR reduction above the national average—or unsuccessful. We next examine the trends of four IMR subgroups—i.e., the PMR (preterm mortality rate): the CMR (congenital malformation-related mortality rate), the SMR (sudden mortality rate), and the OMR (other mortality rate)—in successful and unsuccessful states. Finally, we construct a cross-sectional time series data of the 50 states and run a fixed-effect model to examine the association between the PMR and explanatory variables. Specifically, our study aims to answer the following questions: (1) Which of the four IMR subgroup(s) are most responsible for the reduction of IMR at the national level from 2000 to 2015? (question 1); (2) Which subgroup(s) are most responsible for the IMR reduction in states that have successfully reduced their IMRs from 2000 to 2015? (question 2); and (3) Are variations in teen pregnancy, multiple birth, prenatal care, and multiple births associated with state variations in the PMR? (question 3).

---

[1] Vermont is not shown because of data confidentiality.

## Method

We obtained individual-level data from the linked birth-infant deaths period data files for the periods 2000 to 2015 (14) to compare states' performance in reducing the infant mortality rate and find the factors that might be associated with the PMR. We divided the IMR into four main subgroups: the preterm-related mortality rate (PMR), the congenital malformations-related mortality rate (CMR), the sudden mortality rate (SMR), and other mortality rates (OMR) that includes all other causes of infant mortality. The first subgroup, PMR, is created based on Callaghan and colleagues' criteria (15) and includes those with an underlying cause of death assigned to one of the following ICD-10 categories: K55.0, P010.0, P01.1, P01.5, P020, P02.0, P02.1, P02.7, P07, P10.2, P22, P27, P28.0, P28.1, P36, P52.0, P52.1, P52.2, P52.3, and P77.

The ICD-10 code only considers "Disorders related to short gestation and low birth weight, not elsewhere classified (P07)" as PMR. Callaghan and colleagues showed that the ICD-10 criteria for assigning causes of death do not capture the true number of prematurity-related infant deaths. Callaghan and colleagues categorized preterm-related deaths as those that met two criteria: (1) a biological connection with preterm birth, and (2) more than 75% of infants for such a given cause are born preterm (15).

Congenital malformations, deformations, and chromosomal abnormalities (ICD10: Q00–Q99) are classified under the CMR as the second subgroup. The third subgroup, the SMR, includes infants that died due to sudden infant death (SID) syndrome (ICD10: R95). All other causes are classified as OMR (Appendix A provides a detailed list of the causes and their related ICD10 code). These four categories—the PMR, CMR, SMR, and OMR—are mutually exclusive, which means that they should add up to the total IMR for each state and year.

To identify the subgroup most responsible for the reduction of the IMR at the national level (question 1), we compared changes in the PMR, CMR, SMR, and OMR between 2000 and 2015. We calculated the percentage decline between 2000 and 2015 in each subgroup and compared the rates using the method described by Mathews and colleagues (16).

To investigate the subgroup most responsible for the IMR reduction in states (question 2), we assigned 41 states to one of two groups: successful and unsuccessful. We chose the national IMR reduction size between 2000 and 2015 (i.e., 0.99) as the cutoff point. A state with a reduction size of 0.99 or more was defined as "successful," while all others were "unsuccessful." We calculated the reduction in the PMR, CMR, SMR, and OMR for successful and unsuccessful states. We used a t-test to discern the statistical differences of mean-comparisons among states. To assess the appropriateness of using the pair-wised Student-t test, we used Bartlett's test for homogeneity of variances to compare mean the mortality rates of successful and unsuccessful states. We dropped nine states for the following reasons. First, some states, such as Massachusetts, had a low IMR in 2000, and a minimal opportunity to improve such rate. However, other states, such as Mississippi, experienced a large reduction in their IMR (1.19, a drop from 10.64 to 9.45), but still remained the worst state in the U.S. in the overall rate of infant death in 2015. Second, states in the two groups were not comparable because the mean IMR in 2000 was higher in successful states—7.53 versus 6.63. Nevertheless, we conducted the same analysis using the full sample of states and report the results in Appendix E .

In addition, we constructed a state-level, cross-sectional time series data of 50 states from 2000 to 2015 and used a fixed-effect, multiple linear regression analysis to investigate the association between the PMR and state-level explanatory variables (question 3). We accounted for serial correlation within states by clustering standard errors at the state level. The advantage of the fixed-effect model is that it controls for unobserved factors that are constant over time. Some of the important predictors of birth outcome were unchanged between 2000 and 2015. For example, although the IMR is higher among African-American women, their birth rates did not change in every state from 2000 to 2015. Our model controlled for such factors that remained the same during the study period. We selected our explanatory variables based on evidence in the literature and availability of data in the linked birth-infant deaths period data files. These variables include teen pregnancy, multiple births at infant level, prenatal care, and tobacco use during pregnancy.

Teen pregnancy is defined in each state as the percentage of mothers who were 19 or younger at the time of birth. Multiple births rate is a representation of live births for cases in which a mother delivered more than one baby in the same pregnancy. Tobacco use is the percentage of mothers who reported smoking at any time during pregnancy.

Prenatal care variables indicates the percentage of women in each state who received care based on the Kessner Index (17). This method defines three levels for prenatal care: adequate, intermediate, and inadequate. Since these three variables are perfectly correlated, we dropped the adequate category, and we interpreted the results with respect to it. The linked birth-infant deaths period data reported on these variables until 2002. For the period after 2003, we constructed variables based on Kessner's criteria presented in Appendix C . The Kessner Index has high predictive value for infant mortality and preterm birth (18, 19).

The PMR is reported as deaths per 1,000 live births, and all explanatory variables are in percentages. Thus, to interpret the regression results, we examined how many deaths per 1,000 could be avoided if an explanatory variable is changed by one percentage point (e.g., how many deaths per 1,000 live births would be avoided if we reduced teen pregnancy by one percentage point). Tobacco use and prenatal care were not available for some states between 2011 to 2015. We therefore ran two regressions. First, we regressed the PMR against teen pregnancy and multiple births using a cross-sectional time series of 50 states from 2000 to 2015 that included 800 data points (i.e., 50 states over 16 years). Second, the PMR and all explanatory variables were regressed in a data set of 46 states between 2000 to 2010 (505 data points). The number of observations in the second regression analysis was 505, because some explanatory variables, such as tobacco use in California, is not available (see Appendix D for a complete list of missing values). We also controlled for the effect of year by adding a continuous year variable. Calculations were performed on a Windows OS, using STATA/MP 14.0 software.

# Results

Table 1 shows the national trends for the IMR values and its subgroups in the U.S. between 2000 and 2015. The IMR declined from 6.89 to 5.90 between 2000 and 2015, representing about one fewer death per 1,000 live births. The preterm-related mortality rate (PMR) was the largest subgroup of IMR in 2000, accounting for 2.46 deaths per 1,000 live births. The PMR subgroup was reduced to 2.07 deaths per 1,000 in 2015. In 2000, 1.41 deaths per 1,000 were caused by congenital malformations, which declined to 1.22 deaths per 1000 in 2015. The sudden mortality rate (SMR), the smallest subgroup of the IMR in 2000 and 2015, also declined from 0.62 to 0.39 deaths per 1,000, respectively. Other causes of deaths (OMR), the second largest subgroup, declined from 2.40 to 2.21 deaths per 1,000. The highest reduction, 0.39 deaths per 1,000 live births, occurred in the PMR subgroup. The percentage change in the SMR exceeded the other subgroups.

*Table 1 Summary of infant mortality rates*

| Variable | 2000 | 2015 | Reduction Size | Reduction Rate | P-Value |
|----------|------|------|----------------|----------------|---------|
| IMR | 6.89 | 5.90 | 1.00 | 14% | <0.001 |
| PMR | 2.46 | 2.07 | 0.39 | 16% | <0.001 |
| CMR | 1.41 | 1.22 | 0.19 | 14% | <0.001 |
| SMR | 0.62 | 0.39 | 0.23 | 36% | <0.001 |
| OMR | 2.40 | 2.21 | 0.19 | 8% | <0.001 |

Figure 2 depicts the subgroups' trends in 2000 and 2015 for the two groups of successful and unsuccessful states. Successful states reduced their rates significantly in each of the subgroups (P-value<0.05) while unsuccessful states only reduced their CMR significantly. When it came to comparing the *difference* in reduction sizes, the PMR was the only subgroup with a significantly different reduction in successful states versus unsuccessful. In other words, changes in the CMR, SMR, and OMR in successful states were not statistically different from changes in these subgroups in unsuccessful states.

In 2000, the mean of the IMR in successful states was 7.08, and it significantly declined to 5.47 (P-Value<0.001) in 2015, representing about 1.61 fewer deaths per 1,000 live births. The largest reduction is in the PMR in the successful states, a 0.90 reduction (P-Value <0.001), from 2.63 to 1.73.

The reduction size in the CMR, SMR, and OMR are all significant and equal to 0.22 (from 1.44 to 1.22 with P-Value<0.001), 0.30 (from 0.63 to 0.33 with P-Value<0.001), and 0.18 (2.37 to 2.19 with P-Value=0.022), respectively. Fifty-six percent of the IMR reduction (0.90 out of 1.61) in successful states was due to PMR reduction.

In unsuccessful states, the IMR mean was 7.14 in 2000, which then declined by 0.65 (P-Value<0.001) per 1,000 live births and reached 6.49. A decrease in PMR from 2.48 deaths per 1,000 to 2.34—equivalent to a reduction size of 0.14— was not significant (P-Value=0.076). The CMR declined significantly by 0.27 from 1.53 to 1.26 (P-Value=0.003), the SMR reduced by 0.09 (P-Value=0.080) from 0.65 to 0.56, and the OMR declined by 0.16 (P-Value=0.352) from 2.49 to 2.33 (Figure 2).

The change in the PMR differed significantly between successful and unsuccessful states (P-Value<0.001). Successful states reduced their PMRs by 0.90, while the other group experienced a slight decrease of 0.14 in their PMRs over the period of 2000 to 2015. The reduction for the CMR, SMR, and OMR was not significantly different (P-Values are 0.115, 0.137, 0.157 respectively) between the two groups of states. Successful states reduced their CMRs by 0.22, which is not significantly different (P-Value= 0.115) than the reduction size in unsuccessful states—0.27. The SMR reduction size for successful states was 0.30, which is not significantly larger (P-Value=0.137) than the reduction size of unsuccessful states—0.09. The OMR reduction size was 0.18 in successful versus 0.16 in unsuccessful states, which was not significantly different (P-Value=0.157). Appendix F depicts the successful and unsuccessful states on the U.S. map.

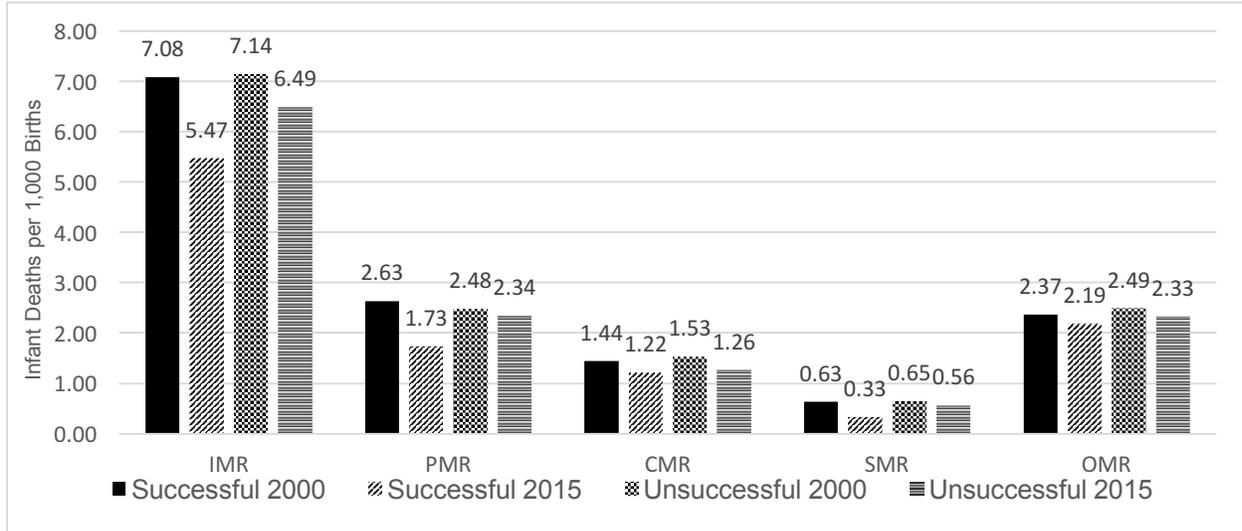

*Figure 2. Performances of successful and unsuccessful states in reduction of IMR and its subgroups in 2000 and 2015 in selected states.*

In summary, our analyses showed that the largest reduction size occurred in the PMR subgroup at the national level. Further, states with an IMR reduction above the national average—successful states—experienced a much larger decline in the PMR subgroups during the period of 2000 to 2015. Successful states had a higher PMR in 2000 compared to unsuccessful states, and reduced it substantially by 2015, while other subgroup reductions were not significantly different between the two categories of states. In the next step, we examine the association between the PMR and some of the determinants of the PMR reported in the literature in a data set of 50 states from 2000 to 2015.

## Regression results for factors associated with PMR

Table 2 shows the summary statistics of each PMR and explanatory variables for selected years of the study period. These covariates are not weighted based on population of states and cannot be interpreted as national estimates. Few states do not report tobacco use, adequate, intermediate, and inadequate care in some years (see Appendix D ). The first column shows the mean of each variable across the entire time frame. On average, 9.34% of pregnant women delivered their infant when they were teenagers, 3.35% of births were multiple births, 12.63% of pregnant women were tobacco users, 68.95% received adequate care, 20.72% received intermediate care, and 9.39% received inadequate care. The percentages of residents with different levels of care did not sum to 100% because of missing values between 2011 to 2015 in some states, such as Connecticut and New Jersey.

*Table 2 Summary Data on Preterm-Related Mortality and State-Level Factors for 50 U.S. States from 2000 to 2015*

| Variable | Mean (SD) | Range | 2000 | 2003 | 2006 | 2009 | 2012 | 2015 |
|---|---|---|---|---|---|---|---|---|
| PMR per 1000 Live Birth | 2.31 (0.74) | 0.52-5.02 | 2.46 | 2.50 | 2.40 | 2.25 | 2.16 | 2.06 |
| % of Teen Pregnancy | 9.34 (2.98) | 3.02-18.75 | 11.80 | 10.27 | 10.20 | 9.92 | 7.76 | 5.86 |
| % of Multiple Births | 3.35 (0.44) | 2.08-4.90 | 3.07 | 3.34 | 3.37 | 3.42 | 3.39 | 3.43 |
| % of Tobacco User | 12.63 (5.04) | 3.84-27.26 | 14.00 | 12.88 | 12.48 | 12.62 | 12.10 | 10.69 |
| % of Adequate Care | 68.95 (8.76) | 39.35-87.38 | 72.86 | 73.31 | 68.93 | 66.86 | 64.92 | 67.76 |
| % of Intermediate Care | 20.72 (5.12) | 9.13-46.95 | 18.08 | 17.98 | 18.31 | 22.11 | 23.76 | 21.50 |
| % of Inadequate Care | 9.39 (5.62) | 1.39-41.00 | 5.07 | 8.71 | 8.93 | 11.04 | 11.32 | 10.73 |

*Source*: Linked Birth/ Infant Death Period Data, 2000-2015
Note 1: Data represented unweighted averages across states, and should not be interpreted as national estimates.
Note 2: Tobacco Use, Adequate, Intermediate, and Inadequate Care have missing data points (see Appendix D )

Table 3 shows the results for the two fixed-effect models. The first regression investigates the association between the PMR and two explanatory variables—teen pregnancy and multiple births— from 2000 to 2015 for 50 states, controlling for year effect and using the state-level fixed effect to control for differences between states. A one-percentage-point increase in the multiple birth rate is significantly associated with 0.18% more deaths per 1,000 live births. Teen pregnancy is not significant. The second regression reports the result when an institutional factor—prenatal care—and a maternal behavior characteristic—tobacco use—are added to the fixed-effect model. Since some states do not report tobacco use and prenatal care after 2010, the data set of the second regression include 46 states from 2000 to 2010. In this regression, the only significant explanatory variable is "inadequate care"—the percentage of pregnant women who do not receive adequate care in a state. A one-percentage-point decline in "inadequate care," from 9.39% (mean reported in Table 2) to 8.39%, is associated with 0.011% less preterm-related deaths per 1,000 live births. The magnitude of this variable is not large relative to the state-level PMR mean (2.31).

*Table* 3 *Parameter Estimates for Fixed-Effect Regression Models of the Preterm-Related Mortality Rate*

| Variables | Regression 1 *(n=800)* | | Regression 2 (n=505) | |
|---|---|---|---|---|
| PMR | Estimate | *P-Value* | Estimate | *P-Value* |
| % of Teen Pregnancy | 0.009 | 0.653 | 0.002 | 0.966 |
| % of Multiple Births | 0.182 | 0.021** | 0.117 | 0.429 |
| % of Tobacco | | | -0.035 | 0.127 |
| % of Intermediate Care | | | 0.007 | 0.315 |
| % of Inadequate Care | | | 0.011 | 0.027** |

'**' Sig at 0.05
Note: Data represented unweighted averages across states, and should not be interpreted as national estimates.
Regression 1 is for 50 U.S. states from 2000 to 2015 and regression 2 is for 46 U.S. states from 2000 to 2010.

## Discussion

Despite the historic declines in the national infant mortality rate (IMR) during the past 16 years, significant variation in the reduction of infant mortality across states still exists. Some states have experienced an increase in their IMRs from 2000 to 2015. We categorized states into successful and unsuccessful groups based on their performance in reducing their IMRs and examined patterns in the leading causes of infant mortality across them. We found that decline in PMR was the major source of success in reducing the IMR during 2000-2015. A decline in the PMR accounts for 0.90 fewer deaths per 1,000—an approximate 56% reduction in the IMR—in successful states.

In addition, we found that preterm-related deaths fell more than other subgroups at the national level. The CDC listed congenital malformation as the first leading cause of infant death in the U.S. from 2005 to 2014 (7) and in 2013 (16). The CDC report, as well as other reports on the IMR (10, 20, 21), only considered short gestation and low birth weight in calculating the PMR. Callaghan and colleagues demonstrated that this classification of infant death does not capture the true magnitude of preterm-related deaths (15). We therefore used the criteria reported by Callaghan and colleagues to identify preterm-related deaths.

Despite statewide variations in the reduction of IMR and PMR, we found a homogenous improvement in the sudden mortality rate (SMR) across states. National-level initiatives such as the "Safe Sleep Campaign," which started in 1994, increased public awareness and led to lower rates of prone infant sleeping (22, 23). The rate of mortality related to sleeping dropped 36% on average in the U.S. between 2000 and 2015. Reduction of the other-related mortality rate (OMR) also did not differ in successful and unsuccessful states from 2000 to 2015. However, the decline in the OMR (8%) was substantially less than the 36% reduction seen in the SMR.

Prenatal care may improve birth outcomes even in low income and resource settings (24-26). Prior studies have found that prenatal care is associated with reduced PMRs. Mothers who did not receive any prenatal care between 2005 and 2009 in Southeastern states, which have the highest rates of infant mortality and lowest average incomes, had the highest IMRs (8). Although we found a significant association between access to prenatal care and the PMR, the magnitude of the coefficient was small. We used a fixed-effect regression analysis to identify variables that might be associated with PMR variations across states. Our analysis revealed that a one-percentage-point decline in mothers that received inadequate care led to a decrease in preterm-related deaths of 0.011 per 1,000 live births. The magnitude of this variable relative to the mean PMR is not large, suggesting that an increase in prenatal care as currently practiced would not lead to a state-level reduction in the PMR. The small-effect size of prenatal care might also be due to a limitation of data collection or lack of control for other key explanatory variables such as adhering to prevention programs.

Despite the emphasis on prenatal care, access to obstetric services is decreasing in the U.S. More than 9% of rural counties lost their obstetric services from 2004 to 2014, which made access to care harder for 28 million women of reproductive age (27). Our findings highlight the importance of eliminating an inadequacy of care. Models such as Coordinated Care Organizations significantly increase access to care with broad financing and delivery reforms that can reduce disparities in prenatal care and improve birth outcomes through a reduction in inadequate care (28). The adoption of expanded access to Medicaid has also been an effective policy. States that adopted Medicaid expansion have observed a reduction in their IMRs (29).

The risk factors, corresponding interventions, and causes of preterm mortality vary by target population. Iams and colleagues classified all possible interventions for a reduction of PMR risk factors into three main categories: primary (directed to all women), secondary (focusing on reducing existing risk), and tertiary (for improving preterm-infant outcomes) (30). Of these, the most improvement in the survival of preterm babies can be attributed to better neonatal care access (4). Obstetric interventions, such as antenatal corticosteroid treatment for women delivering preterm, also contribute to the decline in neonatal mortality.

Another preconceptional strategy to decrease infant mortality is the adoption of steps to reduce the risk of higher-order gestation (4). We found that the percentage of women with multifetal pregnancies is significantly and positively associated with the PMR in the first regression in which we controlled for teen pregnancy. After adding prenatal care variables and tobacco use to our model, multifetal births still correlate positively with the PMR, but statistical significance is lost, perhaps for two reasons. First, the second regression has fewer observations (for 46 states from 2000 to 2010) than the first regression (50 states from 2000 to 2015). Second, the first regression might have omitted the variable bias so that the P-value of multifetal pregnancy changed when we controlled for prenatal care and tobacco use. Multifetal pregnancy substantially increases the risk of preterm birth and infant death (16, 31, 32). In 2013, the IMR for twins (24.37) was more than 4 times the rate for single births (5.25) (16). The IMR for triplets (61.08) was nearly 12 times, and the rate for quadruplets (137.04) was 26 times the rate for single births. However, previous studies (16, 31, 32) did not use a multivariate analysis to disentangle the effects of multiple risk factors over time. Our analysis investigates the percentage impact of multifetal birth at the state level to examine the factors that might be associated with state variations in infant mortality while previous studies compared the IMR between singleton and multifetal births.

 Cessation of cigarette smoking during pregnancy is recommended because it is a prevalent and preventable cause of infant mortality (33-35). It is estimated that 5% of infant mortality and 5.0%–7.3% of preterm-related deaths are attributable to maternal smoking (36, 37). Our findings are not consistent with these studies, which may be attributed to our different design. Traditional studies of smoking in pregnancy have examined cross-sectional data. We assessed the association of the PMR and smoking using cross-sectional time series data. Notably, between 2000 and 2004 in the U.S., smoking among childbearing age women decreased from 25.5% to 21.7%, while preterm birth rates increased from 11.6% to 12.5% (38). One study reported a model for the most important interventions for 39 countries and suggests that smoking cessation had the lowest contribution in reducing the rate of preterm deaths (39).

Our study was limited by the quality of CDC data that may vary by state and hospital over time (40). In addition, due to a lack of information, we could not control for some behavioral risk factors (e.g., drug abuse prevalence and behavioral factors, such as pregnant women's adherence to prevention programs), which may bias our regression results.

Overall, some states performed better in terms of reducing their IMRs over the past decade. Reducing preterm-related death was the biggest factor in states that have improved their infant deaths. It appears that state variations in reducing preterm-related deaths can be partially explained by better access to prenatal care, although the impact size is not large. More qualitative and in-depth analyses are needed to understand why some states have successfully reduced their preterm-related deaths better than others.

## Appendix A  List of OMR Causes

From wonder.cdc.gov:

1) Location: The state, region and division data are derived from the "STRESFIPB" variable in the public use files for years 1999-2002, and from "MRSTATEFIPS" in the public use files for years 2003-2004, and from "MRTERR" for years 2005-2015. The county data are derived from the combined values in variables "STRESFIPB" + "CNTYRFPB" in the public use files for years 1999-2002, and from "MRSTATEFIPS" + "MRCNTYFIPS" in the public use files for years 2003-2004, and from "MRTERR"+ "MRCNTY" for years 2005-2015.

2) Gestational age at birth based on Last Menstrual Period (LMP).

3) The "Gestational Age at Birth" data for years 1995-1998 and the "Gestational Age 10" (formerly named "Gestational Age Group2") data are derived from the "GESTAT10" variable in the public use data for years 1995-2002, and derived from the "GESTREC10" variable in the public use data for years 2003 and later.

A00-B99 (Certain infectious and parasitic diseases)

C00-D48 (Neoplasms)

D50-D89 (Diseases of the blood and blood-forming organs and certain disorders involving the immune mechanism)

E00-E88 (Endocrine, nutritional and metabolic diseases)

F01-F99 (Mental and behavioural disorders)

G00-G98 (Diseases of the nervous system)

H00-H57 (Diseases of the eye and adnexa)

H60-H93 (Diseases of the ear and mastoid process)

I00-I99 (Diseases of the circulatory system)

J00-J98 (Diseases of the respiratory system)

K00-K14 (Diseases of oral cavity, salivary glands and jaws)

K20-K31 (Diseases of oesophagus, stomach and duodenum)

K35-K38 (Diseases of appendix)

K40-K46 (Hernia)

K50-K52 (Noninfective enteritis and colitis)

K55.1 (Chronic vascular disorders of intestine)

K55.2 (Angiodysplasia of colon)

K55.8 (Other vascular disorders of intestine)

K55.9 (Vascular disorder of intestine, unspecified)

K56 (Paralytic ileus and intestinal obstruction without hernia)

K57 (Diverticular disease of intestine)

K58 (Irritable bowel syndrome)

K59 (Other functional intestinal disorders)

K60 (Fissure and fistula of anal and rectal regions)

K61 (Abscess of anal and rectal regions)

K62 (Other diseases of anus and rectum)

K63 (Other diseases of intestine)

K65-K66 (Diseases of peritoneum)

K70-K76 (Diseases of liver)

K80-K86 (Disorders of gallbladder, biliary tract and pancreas)

K90-K92 (Other diseases of the digestive system)

L00-L98 (Diseases of the skin and subcutaneous tissue)

M00-M99 (Diseases of the musculoskeletal system and connective tissue)

N00-N98 (Diseases of the genitourinary system)

O00-O99 (Pregnancy, childbirth and the puerperium)

P00 (Newborn affected by maternal conditions that may be unrelated to present pregnancy)

P01.2 (Newborn affected by oligohydramnios)

P01.3 (Newborn affected by polyhydramnios)

P01.4 (Newborn affected by ectopic pregnancy)

P01.6 (Newborn affected by maternal death)

P01.7 (Newborn affected by malpresentation before labour)

P01.8 (Newborn affected by other maternal complications of pregnancy)

P01.9 (Newborn affected by maternal complication of pregnancy, unspecified)

P02.2 (Newborn affected by other and unspecified morphological and functional abnormalities of placenta)

P02.3 (Newborn affected by placental transfusion syndromes)

P02.4 (Newborn affected by prolapsed cord)

P02.5 (Newborn affected by other compression of umbilical cord)

P02.6 (Newborn affected by other and unspecified conditions of umbilical cord)

P02.8 (Newborn affected by other abnormalities of membranes)

P02.9 (Newborn affected by abnormality of membranes, unspecified)

P03 (Newborn affected by other complications of labour and delivery)

P04 (Newborn affected by noxious influences transmitted via placenta or breast milk)

P05 (Slow fetal growth and fetal malnutrition)

P08 (Disorders related to long gestation and high birth weight)

P10.0 (Subdural haemorrhage due to birth injury)

P10.1 (Cerebral haemorrhage due to birth injury)

P10.3 (Subarachnoid haemorrhage due to birth injury)

P10.4 (Tentorial tear due to birth injury)

P10.8 (Other intracranial lacerations and haemorrhages due to birth injury)

P10.9 (Unspecified intracranial laceration and haemorrhage due to birth injury)

P11 (Other birth injuries to central nervous system)

P12 (Birth injury to scalp)

P13 (Birth injury to skeleton)

P14 (Birth injury to peripheral nervous system)

P15 (Other birth injuries)

P20 (Intrauterine hypoxia)

P21 (Birth asphyxia)

P23 (Congenital pneumonia)

P24 (Neonatal aspiration syndromes)

P25 (Interstitial emphysema and related conditions originating in the perinatal period)

P26 (Pulmonary haemorrhage originating in the perinatal period)

P28.2 (Cyanotic attacks of newborn)

P28.3 (Primary sleep apnoea of newborn)

P28.4 (Other apnoea of newborn)

P28.5 (Respiratory failure of newborn)

P28.8 (Other specified respiratory conditions of newborn)

P28.9 (Respiratory condition of newborn, unspecified)

P29 (Cardiovascular disorders originating in the perinatal period)

P35 (Congenital viral diseases)

P37 (Other congenital infectious and parasitic diseases)

P38 (Omphalitis of newborn with or without mild haemorrhage)

P39 (Other infections specific to the perinatal period)

P50 (Fetal blood loss)

P51 (Umbilical haemorrhage of newborn)

P52.4 (Intracerebral (nontraumatic) haemorrhage of newborn)

P52.5 (Subarachnoid (nontraumatic) haemorrhage of newborn)

P52.6 (Cerebellar (nontraumatic) and posterior fossa haemorrhage of newborn)

P52.8 (Other intracranial (nontraumatic) haemorrhages of newborn)

P52.9 (Intracranial (nontraumatic) haemorrhage of newborn, unspecified)

P53 (Haemorrhagic disease of newborn)

P54 (Other neonatal haemorrhages)

P55 (Haemolytic disease of newborn)

P56 (Hydrops fetalis due to haemolytic disease)

P57 (Kernicterus)

P58 (Neonatal jaundice due to other excessive haemolysis)

P59 (Neonatal jaundice from other and unspecified causes)

P60 (Disseminated intravascular coagulation of newborn)

P61 (Other perinatal haematological disorders)

P70-P74 (Transitory endocrine and metabolic disorders specific to newborn)

P76 (Other intestinal obstruction of newborn)

P78 (Other perinatal digestive system disorders)

P80-P83 (Conditions involving the integument and temperature regulation of newborn)

P90-P96 (Other disorders originating in the perinatal period)

R00-R09 (Symptoms and signs involving the circulatory and respiratory systems)

R10-R19 (Symptoms and signs involving the digestive system and abdomen)

R20-R23 (Symptoms and signs involving the skin and subcutaneous tissue)

R25-R29 (Symptoms and signs involving the nervous and musculoskeletal systems)

R30-R39 (Symptoms and signs involving the urinary system)

R40-R46 (Symptoms and signs involving cognition, perception, emotional State and behaviour)

R47-R49 (Symptoms and signs involving speech and voice)

R50-R68 (General symptoms and signs)

R70-R79 (Abnormal findings on examination of blood, without diagnosis)

R80-R82 (Abnormal findings on examination of urine, without diagnosis)

R83-R89 (Abnormal findings on examination of other body fluids, substances and tissues, without diagnosis)

R90-R94 (Abnormal findings on diagnostic imaging and in function studies, without diagnosis)
R95 (Sudden infant death syndrome)
R99 (Other ill-defined and unspecified causes of mortality)

U00-U99 (Codes for special purposes)

V01-Y89 (External causes of morbidity and mortality)

R96 (Other sudden death, cause unknown)

R98 (Unattended death)

15 *Appendix B  Causes of Mortality per state*

16 Table 4A represents the data for IMR and all four subgroups for each state in 2015.

17 *Table 4A IMR and Subgroups per State*

| No | State | IMR | PMR | CMR | SMR | OMR |
|----|-------|-----|-----|-----|-----|-----|
| 1 | Alabama | 8.31 | 2.77 | 1.46 | 1.07 | 3.02 |
| 2 | Alaska | 6.91 | 1.86 | 1.33 | 0.89 | 2.84 |
| 3 | Arizona | 5.47 | 1.63 | 1.34 | 0.16 | 2.34 |
| 4 | Arkansas | 7.53 | 1.85 | 1.59 | 1.57 | 2.49 |
| 5 | California | 4.43 | 1.57 | 1.12 | 0.24 | 1.50 |
| 6 | Colorado | 4.66 | 1.91 | 1.01 | | 1.71 |
| 7 | Connecticut | 5.65 | 2.83 | 0.56 | | 2.04 |
| 8 | Delaware | 9.22 | 4.66 | 1.07 | | 3.40 |
| 9 | Florida | 6.24 | 2.37 | 1.20 | 0.28 | 2.39 |
| 10 | Georgia | 7.79 | 2.91 | 1.45 | 0.84 | 2.58 |
| 11 | Hawaii | 5.70 | 2.12 | 0.98 | | 2.33 |
| 12 | Idaho | 4.69 | 1.01 | 1.49 | 0.44 | 1.75 |
| 13 | Illinois | 6.00 | 2.44 | 1.11 | 0.09 | 2.35 |
| 14 | Indiana | 7.31 | 2.43 | 1.69 | 0.40 | 2.77 |
| 15 | Iowa | 4.23 | 1.04 | 1.09 | 0.48 | 1.62 |
| 16 | Kansas | 5.95 | 2.12 | 1.46 | 0.46 | 1.92 |
| 17 | Kentucky | 6.68 | 1.79 | 1.38 | 1.04 | 2.50 |
| 18 | Louisiana | 7.56 | 2.57 | 1.21 | 0.70 | 3.09 |
| 19 | Maine | 6.58 | 2.06 | 1.27 | | 2.70 |
| 20 | Maryland | 6.59 | 3.10 | 1.17 | 0.83 | 1.49 |
| 21 | Massachusetts | 4.32 | 2.20 | 0.53 | 0.15 | 1.43 |
| 22 | Michigan | 6.53 | 2.38 | 1.34 | 0.29 | 2.51 |
| 23 | Minnesota | 5.17 | 1.86 | 1.25 | 0.16 | 1.90 |
| 24 | Mississippi | 9.45 | 3.10 | 1.51 | 0.47 | 4.38 |
| 25 | Missouri | 6.49 | 2.08 | 1.43 | 0.24 | 2.73 |
| 26 | Montana | 5.80 | 0.95 | 1.59 | 0.95 | 2.30 |
| 27 | Nebraska | 5.70 | 1.31 | 1.57 | 0.82 | 1.95 |
| 28 | Nevada | 5.18 | 1.29 | 1.35 | | 2.42 |
| 29 | New Hampshire | 4.10 | 1.21 | | | 2.41 |
| 30 | New Jersey | 4.68 | 1.60 | 0.85 | 0.44 | 1.80 |
| 31 | New Mexico | 5.07 | 1.01 | 1.39 | | 2.60 |
| 32 | New York | 4.63 | 1.64 | 0.84 | 0.09 | 2.06 |
| 33 | North Carolina | 7.35 | 2.92 | 1.22 | 0.15 | 3.05 |
| 34 | North Dakota | 7.16 | 1.77 | 1.77 | 1.24 | 2.39 |

| No | State | IMR | PMR | CMR | SMR | OMR |
|---|---|---|---|---|---|---|
| 35 | Ohio | 7.18 | 2.76 | 1.32 | 0.60 | 2.49 |
| 36 | Oklahoma | 7.30 | 2.39 | 1.54 | 0.94 | 2.43 |
| 37 | Oregon | 5.15 | 1.86 | 0.99 | 0.50 | 1.80 |
| 38 | Pennsylvania | 6.15 | 2.55 | 1.00 | 0.47 | 2.14 |
| 39 | Rhode Island | 5.91 | 2.91 | | | 2.00 |
| 40 | South Carolina | 6.91 | 2.53 | 1.24 | 0.43 | 2.72 |
| 41 | South Dakota | 7.30 | 1.78 | 1.54 | 0.81 | 3.16 |
| 42 | Tennessee | 6.94 | 1.90 | 1.44 | 0.24 | 3.37 |
| 43 | Texas | 5.71 | 1.75 | 1.40 | 0.42 | 2.15 |
| 44 | Utah | 5.02 | 1.46 | 1.28 | 0.45 | 1.83 |
| 45 | Vermont | 4.57 | | | | 2.37 |
| 46 | Virginia | 5.90 | 2.09 | 1.25 | 0.57 | 1.99 |
| 47 | Washington | 4.88 | 1.58 | 1.24 | 0.61 | 1.45 |
| 48 | West Virginia | 7.07 | 2.22 | 1.57 | 0.71 | 2.52 |
| 49 | Wisconsin | 5.80 | 2.13 | 1.21 | 0.18 | 2.28 |
| 50 | Wyoming | 4.89 | 1.67 | 1.29 | | 1.80 |

Note that any states with less than 10 deaths in each subgroup is eliminated due to CDC data user agreement.

22    *Appendix C  ADEQUACY OF PRENATAL CARE: KESSNER INDEX*
23

**ADEQUATE***

| Gestation (weeks)**** | Number of Prenatal Visits |
| --- | --- |
| 13 or less AND | 1 or more or not stated |
| 14-17 AND | 2 or more |
| 18-21 AND | 3 or more |
| 22-25 AND | 4 or more |
| 26-29 AND | 5 or more |
| 30-31 AND | 6 or more |
| 32-33 AND | 7 or more |
| 34-35 AND | 8 or more |
| 36 or more  AND | 9 or more |

**INADEQUATE****

| Gestation (weeks)**** | Number of Prenatal Visits |
| --- | --- |
| 14-21*** AND | 0 or not stated |
| 22-29 AND | 1 or less or not stated |
| 30-31 AND | 2 or less or not stated |
| 32-33 AND | 3 or less or not stated |
| 34 or more AND | 4 or less or not stated |

**INTERMEDIATE:** All combinations other than specified above

* In addition to the specified number of visits indicated for adequate care, the Interval to the first prenatal visit has to be 13 weeks or less (first trimester).

** In addition to the specified number of visits indicated for inadequate care, all Women who started their prenatal care during the third trimester (28 weeks or later) are considered inadequate.

*** For this gestation group, care is considered inadequate if the time of the first visit is not Stated.

**** When month and year are specified but day is missing, input 15 for day. Adequacy categories are in accord with recommendations of American College of Obstetrics and Gynecology and the World Health Organization.

25    *Appendix D  List of missing values for each value*
26    Tobacco use and prenatal care have many missing values between 2011-2014. California did not report
27    tobacco use for any year. Therefore, we will remove California in the second regression. Additionally,
28    Florida, Georgia, and Michigan have incomplete or missing values for prenatal care and smoking.
29    Therefore, we removed these States as well.
30    In 2011, 15 areas with "Not Available" data for tobacco use:
31    Alabama", "Alaska", "Arizona", "Arkansas", "Connecticut", "Hawaii", "Maine", "Massachusetts",
32    "Michigan", "Minnesota", "Mississippi", "New Jersey", "Rhode Island", "Virginia", "West Virginia"
33    In 2012, 13 areas with "Not Available" data for tobacco use: "Alabama", "Alaska", "Arizona", "Arkansas",
34    "Connecticut", "Hawaii", "Maine", "Michigan", "Mississippi", "New Jersey", "Rhode Island", "Virginia",
35    "West Virginia"
36    In 2013, 10 areas with "Not Available" data for tobacco use: "Alabama", "Arizona", "Arkansas",
37    "Connecticut", "Hawaii", "Maine", "Michigan", "New Jersey", "Rhode Island", "West Virginia"
38    In 2014, 4 areas with "Not Available" data for tobacco use: "Connecticut", "Hawaii", "New Jersey", "Rhode
39    Island"
40    In 2015, 2 areas with "Not Available" data for tobacco use: "Connecticut", "New Jersey"
41
42
43

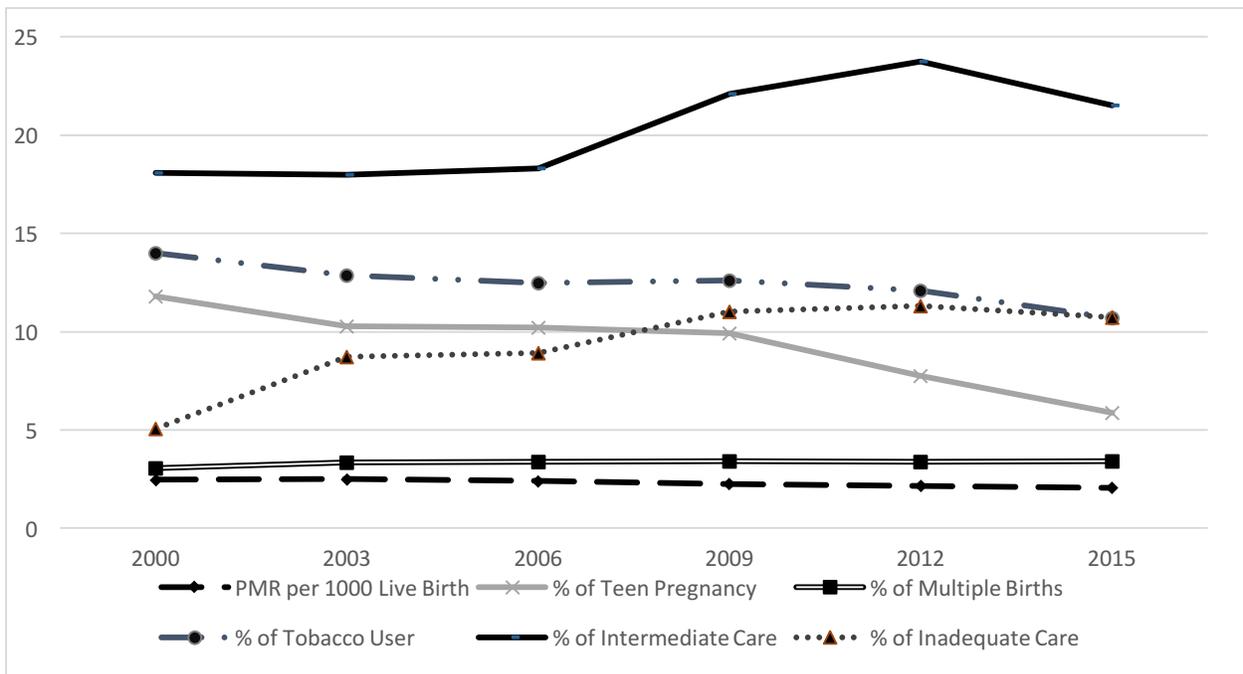

44
45    *Figure 3A Data on Preterm-Related Mortality and State-Level Factors for 50 U.S. States*

46

47 *Appendix E  Performance Analysis of full list of States*

48       Without dropping any states from categories, the mean IMR in successful states was 7.53 in 2000

49 and declined to 5.90 in 2015 (P-Value<0.001), representing 1.63 less deaths per 1,000 live births (Figure

50 4A). In successful states, reduction in the PMR accounted for the largest reduction in IMR—a 0.76

51 reduction (P-Value <0.001) from 2.63 to 1.87 (Figure 4A). Forty-seven percent of the IMR reduction (0.76

52 out of 1.63) in successful states was due to the PMR reduction. The reduction size in the CMR, SMR, and

53 OMR were 0.32 (from 1.57 to 1.25 with P-Value=0.001), 0.34 (from 0.72 to 0.38 with P-Value<0.001), and

54 0.20 (2.60 to 2.40 with P-Value=0.086), respectively.

55       The IMR mean in unsuccessful states was 6.63 in 2000, which then declined by 0.29 per 1,000

56 live births to 6.34 deaths per 1,000 in 2015 (P-Value=0.051). In unsuccessful states, the PMR slightly

57 decreased by 0.04 from 2.28 deaths per 1,000 to 2.24 (P-Value=0.60), the CMR declined by 0.12 from

58 1.37 to 1.25 (P-Value=0.017), the SMR reduced by 0.13 from 0.68 to 0.55 (P-Value=0.069), and the OMR

59 stayed almost the same by 0.01 decline from 2.30 to 2.29 (P-Value=0.967) (Figure 4A).

60       The changes in the PMR differed significantly between successful and unsuccessful states (P-

61 Value<0.001). Successful states reduced their PMRs by 0.76, while the other group experienced a slight

62 decrease in the PMR—0.04— over the period of 2000 to 2015. The reduction for the CMR and SMR was

63 also significantly different between the two groups of states (P-Values of 0.028 and 0.036, respectively).

64 Successful states reduced their CMRs by 0.32, which is more than 2.5 times of the reduction size in

65 unsuccessful states—0.12. The SMR reduction size for successful states was 0.34, which is significantly

66 larger than the reduction size of unsuccessful states—0.13. Although the OMR size was the largest after

67 the PMR, its reduction size was not large in any categories of states (0.20 in successful versus 0.01 in

68 unsuccessful states), this difference was not significant (P-Value=0.172).

69

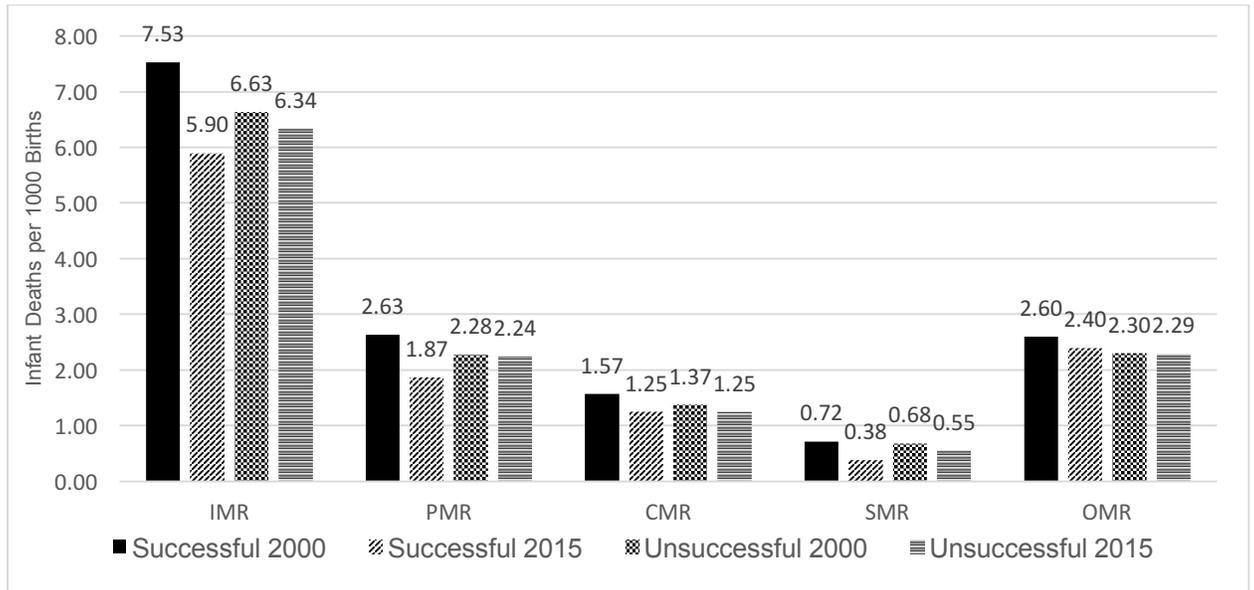

70

*Figure 4A Performances of successful and unsuccessful states in reduction of IMR and its subgroups in 2000 and 2015*

72

73 *Appendix F  Category of states*
74 Table 5A List of States Based on Performance in Reduction of IMR between 2000 to 2015

| Successful States | | Unsuccessful States | | States dropped from the sample | |
|---|---|---|---|---|---|
| state | Reduction Size | state | Reduction Size | state | Reduction Size |
| Arizona | -1.28 | Alaska | 0.00 | Alabama | -1.20 |
| California | -0.99 | Arkansas | -0.62 | Louisiana | -1.47 |
| Colorado | -1.49 | Connecticut | -0.86 | Maine | 1.73 |
| Hawaii | -2.39 | Delaware | -0.37 | Massachusetts | -0.27 |
| Idaho | -2.87 | Florida | -0.67 | Mississippi | -1.19 |
| Illinois | -2.49 | Georgia | -0.66 | South Dakota | 2.08 |
| Iowa | -2.20 | Indiana | -0.46 | Tennessee | -2.13 |
| Michigan | -1.67 | Kansas | -0.68 | Utah | -0.24 |
| Nebraska | -1.44 | Kentucky | -0.42 | Washington | -0.32 |
| Nevada | -1.34 | Maryland | -0.93 | | |
| New Hampshire | -1.72 | Minnesota | -0.45 | | |
| New Jersey | -1.59 | Missouri | -0.70 | | |
| New Mexico | -1.54 | Montana | -0.13 | | |
| New York | -1.77 | Ohio | -0.50 | | |
| North Carolina | -1.27 | Oregon | -0.42 | | |
| North Dakota | -1.18 | Pennsylvania | -0.96 | | |
| Oklahoma | -1.17 | Rhode Island | -0.40 | | |
| South Carolina | -1.85 | Texas | 0.09 | | |
| Vermont | -1.73 | West Virginia | -0.36 | | |
| Virginia | -1.01 | Wisconsin | -0.83 | | |
| Wyoming | -1.82 | | | | |

75
76 To make initial average rates of IMR comparable among two groups of successful and unsuccessful
77 states, we removed states with high initial IMR (IMR at 2000 > 9) from "Successful" and states with low
78 initial IMR (IMR at 2000 < 5.5) from "Unsuccessful" and repeated the analysis. Therefore, we removed
79 Alabama, Louisiana, Mississippi, and Tennessee from successful states because of high initial IMR in
80 2000—9.51, 9.03, 10.64, and 9.07 respectively. Also, we dropped Massachusetts, Maine, South Dakota,
81 Utah, and Washington with low initial IMR from unsuccessful states— 4.59, 4.85, 5.22, 5.26, and 5.20
82 respectively (The third column in Table 4).**Error! Reference source not found.**
83

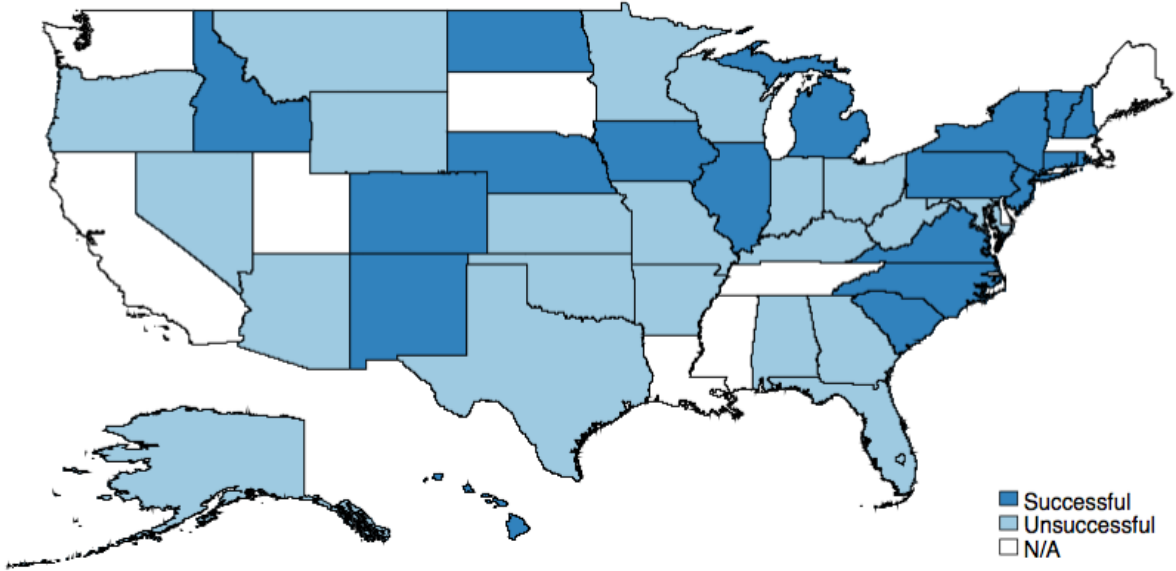

*Figure 5A Category of state Based on Performance in Reduction of IMR between 2000 and 2015*

84
85

86